\documentstyle[prd,aps,preprint,eqsecnum]{revtex}
\begin{document}
\preprint{SUSSEX-AST 94/4-1\\gr-qc/9405025}
\draft
\tighten
\title{Massless fields in scalar-tensor cosmologies}
\author{Jos\'{e} P.~Mimoso\cite{present} \& David Wands}
\address{Astronomy Centre, School of Mathematical and Physical Sciences, \\
University of Sussex, Brighton BN1 9QH.~~~U.~K.}
\date{\today}
\maketitle
\begin{abstract}
We derive exact Friedmann--Robertson--Walker cosmological solutions in general
scalar--tensor gravity theories, including Brans--Dicke gravity, for stiff
matter or radiation. These correspond to the long or short wavelength modes
respectively of massless scalar fields. If present, the long wavelength modes
of such fields would be expected to dominate the energy density of the
universe at early times and thus these models provide an insight into the
classical behaviour of these scalar--tensor cosmologies near an initial
singularity, or bounce. The particularly simple exact solutions also provide a
useful example of the possible evolution of the Brans--Dicke (or dilaton)
field, $\phi$, and the Brans--Dicke parameter, $\omega(\phi)$, at late times
in spatially curved as well as flat universes. We also discuss the
corresponding solutions in the conformally related Einstein metric.
\end{abstract}



\section{Introduction}

If we hope to describe gravitational interactions at energy densities
approaching the Planck scale it seems likely that we will need to
consider Lagrangians extended beyond the Einstein--Hilbert
action of general relativity.  The low--energy effective action in
string theory, for instance, involves a dilaton field coupled to the
Ricci curvature tensor\cite{string}.  Scalar fields coupled directly to the
curvature appear in all dimensionally reduced gravity theories, and their
influence on cosmological models was first seriously considered by
Jordan\cite{Jordan 59}. These models have been termed scalar--tensor gravity,
the best known of these being the Brans--Dicke theory\cite{Brans+Dicke 61}.
Gravity Lagrangians including terms of higher order in the Ricci scalar can
also be cast as scalar--tensor theories\cite{Tey+Tou 83,Wands 94} with
appropriate scalar potentials.

The belief that modified gravity theories may have played a crucial role
during the early universe has recently been rekindled by extended
inflation\cite{ExtInf}. In this scenario
a scalar--tensor gravity theory
allows the first order phase transition of the ``old'' inflationary
model\cite{Guth 81} to complete. This arises because the scalar field,
$\phi$, (henceforth the Brans--Dicke field, essentially the inverse of
the Newton's gravitational ``constant'') damps the rate of expansion and,
in the original extended inflationary scenario based
on the Brans--Dicke theory, turns the exponential expansion found in
general relativity into
power law inflation\cite{powerlaw}. However, Brans--Dicke theory is
unable to meet the simultaneous and disparate requirements placed by the
post--Newtonian solar system tests\cite{Reas+al79} and by the need to
keep the sizes of the bubbles nucleated during inflation within the
limits  permitted by the anisotropies of the microwave
background\cite{buts}.

This situation may be averted through the consideration of more general
scalar--tensor  theories in which the parameter $\omega$, a constant in
Brans--Dicke theory, is allowed to vary as $\omega(\phi)$~\cite{HyperInf 90}.
But in such cases one needs to understand better the  cosmological behaviour
of these general theories, in order to assess their implications on our
models of the universe.  Direct observations mainly constrain these
theories at the present day in our solar system\cite{Will 92}  imposing
a lower bound $\omega >500$ and requiring that $\omega^{-3}
(d\omega/d\phi)$ should approach zero. On a cosmological scale,
the principal limits arise from the  consideration of effects upon the
synthesis of light elements, indicating that at the time of
nucleosynthesis similar bounds hold\cite{nuc}.

Most of the work that can be found in the literature on solutions of
scalar--tensor theories concerns the particular case of Brans--Dicke
theory\cite{Nariai 68,O-H+Tup 72,Gur+Fin+Rub 73,L-P,Weinberg 72,Morganstern
71}. The properties of more general scalar-tensor cosmologies have been
discussed recently\cite{Damour+Nordvedt 93,BMM 93} and exact solutions derived
for the vacuum and radiation models\cite{Barrow 93} (where $p=\rho/3$)
corresponding to the particular situation where the scalar field is sourceless
(because the matter energy--momentum tensor is traceless). In this paper we
show how to extend this method to derive exact solutions for the homogeneous
and isotropic cosmological models with a perfect fluid characterized by the
equation of state $p=\rho$, which does act as a source for the Brans--Dicke
field. These models represent the evolution a homogeneous massless scalar
field\cite{Tab+Tau 73}. Such a scalar field may describe the evolution of
effectively massless fields, including in the context of superstring cosmology
the antisymmetric tensor field which appears in the low energy string
effective action\cite{string,CopLahWan 94}.

As the energy density of a barotropic perfect fluid with $p=(\gamma-1)\rho$
evolves as $\rho\propto a^{-3\gamma}$, such ``stiff matter'' would be expected
to dominate at early times in the universe\cite{Barrow 78} over
short wavelength modes or any other matter with $p<\rho$. Thus our solutions
provide an important indication of the possible early evolution of
scalar--tensor cosmologies.

As in~\cite{Barrow 93}, our solutions will be given in closed form in
terms of an integration depending on $\omega(\phi)$, which can be performed
exactly in many cases and numerically in all cases.  The general field
equations are given in Section~\ref{sectstt} for scalar--tensor gravity and we
solve these in Friedmann--Robertson--Walker metrics for vacuum, stiff fluid and
radiation models in Section~\ref{sectFRW}. The equivalent picture in the
conformally transformed Einstein frame is presented in Section~\ref{sectCONF}.
Conclusions on the general behaviour of solutions are presented in our final
section.

\section{Scalar--tensor gravity theories}
\label{sectstt}

The scalar--tensor field equations~\cite{STT} are derived
from the action\footnote
{One can also include an additional
boundary term dependent on the extrinsic curvature of the
boundary~\cite{Wands 94}, as is required in general
relativity~\cite{bcs} to allow for the variation of $g_{ab,c}$ on the
boundary.}
\begin{equation}
S = \frac{1}{16\pi} \int_M d^4x \sqrt{-g} \left[
	\phi \left( R -2\lambda(\phi) \right)
	- \frac{\omega(\phi)}{\phi} g^{ab} \phi_a \phi_b
	+ 16\pi {\cal L}_m \right] \; ,
\label{eSTTaction}
\end{equation}
where $R$ is the usual Ricci curvature scalar of the spacetime,  $\phi$
is the Brans--Dicke scalar field, $\omega(\phi)$ is a dimensionless
coupling function and ${\cal L}_m$ represents the  Lagrangian for the
matter fields. It is clear that the scalar field plays
the role which in general relativity is played by the gravitational
constant, but with $\phi$ now a dynamical variable.

The particular case of Brans--Dicke gravity arises when we take $\omega$
to be a constant and $\lambda=0$ in the Lagrangian of Eq.~(\ref{eSTTaction}).
The $\lambda(\phi)$ potential is the natural generalisation of the
cosmological constant $\Lambda$. It introduces terms which violate
Newtonian gravity at some length scale. In what follows we will leave
$\omega(\phi)$ as a free function but consider only models in which
$\lambda$ is zero. This should be valid at least at sufficiently early
times when we expect kinetic terms to dominate, and will also
avoid introducing too many free functions into our analysis. (Note that the
Lagrangian is sometimes written in terms of a scalar field $\varphi$ with a
canonical kinetic term so that $\phi\equiv f(\varphi)$ and $\omega(\phi)\equiv
f/2(df/d\varphi)^2$.)

Taking the variational
derivatives of the action (\ref{eSTTaction}) with respect to
the two dynamical variables $g_{ab}$ and $\phi$ and setting $\lambda(\phi)=0$
yields the field equations
\begin{eqnarray}
R_{ab}-\frac{1}{2}\,g_{ab}\, R & = &
 8\pi \, {T_{ab}\over \phi} +
 \frac{\omega(\phi)}{\phi^2} \left( g_a^c g_b^d - \frac{1}{2} g_{ab} g^{cd}
				\right) \phi_{,c}\phi_{,d} \nonumber \\
& & \quad + \frac{1}{\phi} \left( \nabla_a\nabla_b\phi - g_{ab} \Box\phi
\right)
\; ,
\label{eSTTFE1}\\
\Box{\phi} & = &
 \frac{1}{2\omega(\phi)+3}\left[ 8\pi \, T - g^{cd} \omega_{,c}\phi_{,d}
 \right]
\label{eSTTFE2}
\; ,
\end{eqnarray}
where $T=T_a^a$ is the trace of the energy--momentum tensor of the
matter defined as
\begin{equation}
T^{ab} = \frac{2}{\sqrt{-g}} \frac{\partial}{\partial g_{ab}}
	\left( \sqrt{-g} {\cal L}_m \right) \; ,
\end{equation}

It is important to notice that the usual  relation $\nabla_b T^{ab}=0\,$
establishing the conservation laws satisfied by the matter fields holds true.
This follows from the assumption that all matter fields are minimally coupled
to the metric $g_{ab}$ which means that the principle of equivalence is
guaranteed. The role of the scalar field is then that of determining the
spacetime curvature (associated with the metric) produced by the matter.
Matter may be a source of the Brans--Dicke field, but the latter acts back on
the matter only through the metric~\cite{Thorne+Will 71}.

\section{Friedmann--Robertson--Walker models}
\label{sectFRW}

We consider homogeneous and isotropic universes with the metric given by the
usual Friedmann--Robertson--Walker (FRW) line element
\begin{equation}
ds^2=-dt^2 + a(t)^2 \left[\frac{dr^2}{1-kr^2} + r^2(d\theta^2 +
\sin{\theta}^2 d\varphi^2)\right] \; .
\end{equation}

The field equations for a scalar--tensor theory, where we allow
the coupling parameter $\omega$ to depend on the scalar field
$\phi$, but restrict the potential $\lambda$ to be zero, are then
\begin{equation}
H^2 +H\, \frac{\dot{\phi}}{ \phi} -\frac{\omega(\phi)}{ 6}\,
\frac{\dot{\phi}^2}{ \phi^2}
+\frac{k}{ a^2} =  \frac{8\pi}{3} \frac{\rho}{ \phi}
\label{eFr} \; ,
\end{equation}
\begin{equation}
\ddot{\phi}+\left[ 3\frac{\dot{a}}{ a}+
\frac{\dot{\omega}(\phi)}{2\omega(\phi)+3}\right]\,\dot{\phi}
= \frac{8\pi\rho}{2\omega(\phi)+3}\; (4-3\gamma)
\label{eSc} \; ,
\end{equation}
\begin{equation}
\dot{H} +H^2+\frac{\omega(\phi)}{ 3} \frac{\dot{\phi}^2}{ \phi^2}-
H\,\frac{\dot{\phi}}{ \phi} = - \frac{8\pi\rho}{3\phi}\;
\frac{(3\gamma-2)\,\omega+3}{2\omega(\phi)+3}+
\frac{1}{2}\,\frac{\dot{\omega}}{2\omega(\phi)+3}\,
\frac{\dot{\phi}}{\phi}
\label{eRay} \; .
\end{equation}
These equations differ from the corresponding equations of
Brans--Dicke theory through  the presence of terms
involving $\dot{\omega}$ in the two latter equations.

Several authors have studied cosmological solutions of the Brans--Dicke theory
for a FRW universes filled with a perfect
fluid~\cite{Brans+Dicke 61,Nariai 68,O-H+Tup 72,Gur+Fin+Rub 73,L-P}.
Nariai~\cite{Nar 68} derived power law solutions
for the flat FRW universe with a perfect fluid satisfying the barotropic
equation of state $p=(\gamma -1) \rho$, with $\gamma$ a constant taking
values in the interval  $0\le \gamma \le 2$.

The solutions of these equations of motion
are defined by four integration constants whereas the
corresponding solutions in general relativity depend on only
three~\cite{Weinberg 72}. In fact, in addition to the values of
$a(t_0)$, $\dot{a}(t_0)$ and $\phi(t_0)\propto 1/G(t_0)\,$,
we now need $\rho(t_0)$ (or $\dot{\phi}(t_0)$ instead) as well. Originally,
as done by Brans and Dicke, and by Nariai,
this extra freedom was eliminated by requiring that
$\dot{\phi} a^3$ should vanish when $a$ approaches
the initial singularity at $a=0$. In a flat FRW model
this restricts one to obtaining only the power law solutions of
the Brans--Dicke theory. Here we shall keep our analysis more general.

The derivation of the general barotropic Brans--Dicke solutions
for the spatially flat ($k=0$) model was done by
Gurevich, Finkelstein and Ruban~\cite{Gur+Fin+Rub 73}.
Their solutions, which cover all the space of parameters of the theory,
render clear a very important feature of the behaviour of the flat
cosmological models, namely that for the solutions which exhibit
an initial singularity (those with
$\omega > -3/2\,$) the scalar field dominates the expansion at early times,
whilst the later stages are matter dominated and approach the behaviour
of Nariai's solutions for $\omega>2(\gamma-5/3)/(2-\gamma)^2$. Note
therefore that the stiff fluid solutions are unique in that they do not
approach Nariai's power law solutions at late times other than in the limit
$\omega\to\infty$.

The same solutions for the flat model in the cases of vacuum, stiff matter
($p=\rho\,$) and radiation ($p=\rho/3\,$) were rederived later by
Lorentz--Petzold~\cite{L-P} using a different method which enabled
him to also obtain solutions for the non--flat models. We use this method
in an improved form~\cite{Barrow 93} to derive solutions for the general
scalar--tensor theories.

Using the conformal time variable $\eta$ defined by the differential relation
\begin{equation}
dt= a\; d\eta \; ,
\end{equation}
and the variables
\begin{equation}
X \equiv \phi\;a^2 \; ,
\end{equation}
and
\begin{equation}
Y \equiv  \int\, \sqrt{\frac{2\omega+3}{3}}\;
\frac{d\phi}{\phi} \; ,
\label{eYphi}
\end{equation}
we can re--write the above field equations as
\begin{equation}
(X')^2 + 4 \, k \,X^2 - (Y'\, X)^2 = 4 M \;X \, a^{4-3\gamma}
\label{eLPam} \; ,
\end{equation}
\begin{equation}
\left[Y'\,X\right]'=M(4-3\gamma)\;\sqrt{\frac{3}{2\omega+3}}\; a^{4-3\gamma}
\label{eLPbm} \; ,
\end{equation}
\begin{equation}
X'' + 4\,k \; X= 3\,M(2-\gamma)\; a^{4-3\gamma}
\label{eLPcm} \; ,
\end{equation}
where the density $\rho=3M/8\pi a^{3\gamma}$ for a barotropic fluid with $M$ a
constant. The prime denotes differentiation with respect to $\eta$. Our
variables are akin to those used by Lorentz--Petzold~\cite{L-P} when solving
for the Brans--Dicke theory. To that extent the method we explore here is a
generalization of his method of obtaining decoupled equations. Note that
whenever $X$ is negative this must correspond to a negative value for $\phi$.
In what follows, unless otherwise explicitly stated, we shall assume that
$\omega>-3/2$ to guarantee the positiveness of the function under the square
root in Eq.~(\ref{eYphi}), although it would be straightforward to redefine
$Y(\phi)$ for the case of $\omega<-3/2$.

This system considerably simplifies for
the two particular cases: $\gamma=4/3$ (radiation) and
$\gamma=2$ (stiff matter). We shall show in the next section
precisely how these correspond to the short and long wavelength limits
of a massless field.
In either case the full integration is again possible, provided we
specify the function $\omega(\phi)$. For
other values of $\gamma$, the equations carry an explicit
dependence on $a$ which cannot be integrated by the method adopted here.
An alternative approach for these latter cases based on another method
of integration of the original field equations is presented
elsewhere~\cite{Barrow+Mimoso 93}.

Anisotropic cosmologies have also been considered in the literature, again
principally for Brans--Dicke gravity. We will show elsewhere how our method
may be extended to derive solutions for general scalar--tensor gravity
in anisotropic models\cite{Mimoso+Wands 94}.


\subsection{Scalar field evolution}

A minimally coupled scalar field, $\sigma$, whose energy--momentum tensor
\begin{eqnarray}
T_{ab} & = & \left( g_a^c g_b^d - \frac{1}{2}g_{ab}g^{cd} \right)
	\sigma_{,c} \sigma_{,d} \nonumber \; ,\\
 & = & (p+\rho) \, u_a u_b + p \, g_{ab} \; ,
\end{eqnarray}
corresponds to a perfect fluid with density
\begin{equation}
\rho = p = {1 \over 2} |g^{ab}\sigma_{,a}\sigma_{,b}| \; ,
\end{equation}
and normalised velocity field
\begin{equation}
u_a = \frac{\sigma_{,a}}{|g^{cd}\sigma_{,c}\sigma_{,d}|^{1/2}} \; .
\end{equation}

The scalar field itself obeys the wave equation
\begin{equation}
\Box \sigma \ = \ 0 \; ,
\end{equation}
which in a FRW metric reduces to
\begin{equation}
- \ddot{\sigma} - 3 H \dot{\sigma}
 + \sum_{i,j=1}^{3} g^{ij} \nabla_i \nabla_j \sigma \ = \ 0 \; .
\end{equation}

If we consider plane wave solutions of the form
$\sigma=\sigma_q(t)\exp(i\sum q_ix^i)$
then for $\left(q/a\right)^2\gg H^2,\dot{H},k/a^2$ this can be re--written as
\begin{equation}
(a\sigma_q)'' + q^2 (a\sigma_q) \ = \ 0 \; ,
\label{esigmarad}
\end{equation}
where $q^2=\sum q_i^2$. This corresponds to the usual flat space result for
plane waves where $(a\sigma_q)\propto\exp(iq\eta)$ and thus
$T_a^b=q_aq^b\sigma^2\propto a^{-4}$ with the null four--vector $q_a=(q,q_i)$.
This is an anisotropic stiff fluid but if we consider an isotropic
distribution of short wavelength modes averaged over all spatial directions
this produces a perfect fluid with $\langle p\rangle=\rho/3$, the usual result
for isotropic radiation.

For long wavelength modes in an FRW universe
(with comoving wavenumber $\left(q/a\right)^2\ll H^2,\dot{H}$) we can
neglect spatial gradients in the field and the first integral of
Eq.~(\ref{esigmarad}) yields $a^3\dot{\sigma}=$constant and thus
$p=\rho=\dot{\sigma}^2/2\propto a^{-6}$, i.e.~a stiff fluid.

Clearly the dividing line between these long and short wavelength modes
changes as the comoving Hubble length or curvature scale
evolves\footnote{Indeed this is precisely how long wavelength perturbations
are produced in the inflaton field from originally short wavelength vacuum
flutuations as the comoving Hubble length shrinks during
inflation\cite{QPert}. This highlights the potential importance of quantum
effects which we shall neglect in this purely classical treatment.}. In a
conventional (non--inflationary) cosmology the comoving Hubble length shrinks
as we consider earlier and earlier times in an expanding universe so that as
$a\to0$ all modes must be ``outside the horizon'' and evolve as a homogeneous
stiff fluid lending support to our contention that the stiff fluid solutions
will be important in determining the classical behaviour of scalar--tensor
cosmologies near any initial singularity. In any case, as already remarked,
the energy density of a barotropic perfect fluid evolves as $\rho\propto
a^{-3\gamma}$ and so the energy density of a stiff fluid will eventually
dominate as $a\to0$ over any matter with a barotropic index $\gamma<2$.

In what follows we shall consider only the extreme short and long wavelength
modes of the massless field, neglecting the intermediate regimes.

\subsection{Vacuum solutions}
\label{sVac}

Let us first consider the field equations in vacuum;
\begin{eqnarray}
(X')^2 - (Y'\, X)^2 + 4 \, k \,X^2 = 0
\label{eLPa} \; , \\
\left( Y'\; X\right)'=0
\label{eLPb} \; , \\
X'' + 4\,k\;X = 0
\label{eLPc} \; .
\end{eqnarray}
Both Eqs.~(\ref{eLPb}) \&~(\ref{eLPc}) are easily integrable, and $X(\eta)$ is
independent of the particular $\omega(\phi)$ dependence.

Solving Eq.~(\ref{eLPc}) yields
\begin{equation}
X(\eta) = \left\{
\begin{array}{cc}
A\; \eta &\qquad {\rm for} \, k=0
\; , \\
\frac{A }{2}\;\sin{(2\eta)} & \qquad {\rm for} \, k=+1
\; , \\
 \frac{A }{2}\;\sinh{(2\eta)} & \qquad {\rm for} \, k=-1
\; ,
\end{array}
\right. \label{eg0}
\end{equation}
with $A$ an arbitrary integration constant [see Fig.~\ref{fig1}].
In what follows we will find it
most useful, and succinct, to write this as
\begin{equation}
X(\eta) = \frac{\pm A\tau}{1+k\tau^2} \; ,
\label{evacX}
\end{equation}
in terms of the new time variable
\begin{equation}
\tau(\eta) =
\left\{
\begin{array}{cc}
	| \eta | & \qquad {\rm for} \, k=0 \; , \\
	| \tan{\eta} | & \qquad {\rm for} \, k=+1 \; ,\\
	| \tanh{\eta} | & \qquad {\rm for} \, k=-1 \; .
\label{etau}
\end{array}
\right.
\end{equation}
It is convenient to define $\tau$ as a non--negative quantity and choose the
plus or minus sign in Eq.~(\ref{evacX}) according to whether $\eta$ is greater
or less than zero respectively. This only amounts to a different choice of the
integration constant and so can be absorbed in our choice of $A$. In practice,
because $a^2=X/\phi$ must always be non--negative only one choice of $\pm A$
corresponds to a real solution anyway. For the allowed choice of $A$, $\tau$
may then either increase or decrease with conformal (and thus also with
proper) time.

{}From Eq.~(\ref{eLPb}) we obtain
\begin{equation}
Y' \; X=  f ={\rm const} \; .
\label{eY'vac}
\end{equation}
This latter result implies
\begin{equation}
Y \equiv \int\,\sqrt{\frac{2\omega+3}{3}}\;\frac{{\rm d}\phi}{\phi} =
\int\,\frac{f}{X}\,{\rm d}\eta \; .
\label{eYvac}
\end{equation}

Now, notice that Eq.~(\ref{eLPc}) has the first integral
\begin{equation}
X'^2+4\,k\;X^2= \bar{A} \; ,
\label{eXint}
\end{equation}
where $\bar{A}=A^2$ for the solutions given in Eq.~(\ref{evacX}).
Thus, substituting Eq.~(\ref{eY'vac}) into Eq.~(\ref{eLPa}) we obtain a
relation between the constants $f$ and $A$ such that $A=\pm f$.

Given the definition of $X$, and the fact that we know $X(\eta)$ from
Eq.~(\ref{eg0}), we realize that provided we know
the particular form of $\omega(\phi)$ we can obtain
$\phi(\eta)$ from Eq.~(\ref{eYvac}), and then derive the scale factor
$a(\eta)$ as
\begin{equation}
a^2(\eta)= \frac{1}{\phi} X \; .
\end{equation}
To obtain $\phi\,$ we have to invert $Y(\phi)$, given by the
left--hand side of Eq.~(\ref{eYvac}) and use the fact that
we know the right--hand side
\begin{equation}
Y(\eta) = \int\,\frac{f}{X}\, d\eta \, = \pm \ln{\tau(\eta)}
\ + \ {\rm constant} \; .
\end{equation}
We see that as $\tau\to0$ or $\tau\to\infty$ (at early or late times) the
function $Y$ must diverge.
For instance in the case of Brans--Dicke gravity where $\omega$ is a constant
this implies that $\phi\to0$ or $\phi\to\infty$.
However, there is {\em a priori} no prescription for $\omega(\phi)$.
Thus, we are led to consider some specific $\omega(\phi)$ dependences
which hopefully will shed some light onto general results
concerning the dependence of the solutions on the form of
$\omega(\phi)$.

Even without solving these equations for a particular $\omega(\phi)$ we can
come to some general conclusions about how these vacuum solutions behave.
As $X\to0$ and $(X'/X)^2\to\infty$ the curvature becomes negligible in
Eq.~(\ref{eLPa}) and we see that
\begin{equation}
\left( \frac{X'}{X} \right)^2 \to \left( Y' \right)^2 \; .
\end{equation}
Thus using the Eq.~(\ref{eYphi}) we have
\begin{eqnarray}
\dot{a} & = & \frac{1}{2} \left( \frac{X'}{X} - \frac{\phi'}{\phi} \right) \; ,
\\
& \to & \frac{1}{2} \left( 1 \mp \sqrt{\frac{3}{2\omega+3}} \right)
  \frac{X'}{X}
\; ,
\end{eqnarray}
and the initial singularity (with $\dot{a}\to\pm\infty$) can only be avoided
for
$\omega\to0$.

A necessary condition for any turning point in the evolution of the scale
factor is
\begin{equation}
\omega(\phi) \ = \ \frac{6kX^2}{A^2-4kX^2} \; .
\end{equation}
Thus for $k\leq0$ a turning point can only occur when $\omega\leq0$. This
corresponds to the $(\nabla\phi)^2$ term in the action of
Eq.~(\ref{eSTTaction}) having the ``wrong sign'', in that it can contribute a
negative effective energy density. Turning points can occur in closed models
even if $\omega>0$, just as they can occur in general relativity. Note that
the sign of the gravitational coupling, $\phi$, (and thus $X$) is irrelevant
in this vacuum case.

\subsubsection*{Vacuum solutions in Brans--Dicke gravity}

Let us consider first the case $\omega(\phi)=\omega_0=$constant,
corresponding to the Brans--Dicke theory. Then
\begin{equation}
Y=\sqrt{\frac{2\omega_0+3}{3}}\; \ln{\frac{\phi}{\phi_0}} \; ,
\end{equation}
and thus
\begin{eqnarray}
\phi & = & \phi_0 \tau^{\pm\beta} \; , \\
a^2 & = & \frac{A}{\phi_0} \frac{\tau^{1\mp\beta}}{1+k\tau^2} \; ,
\end{eqnarray}
where we have written $\beta=\sqrt{3/(2\omega_0+3)}$. These solutions are
plotted in Fig.~\ref{fig2} and Fig.~\ref{fig3}.

The $k=0$ solutions correspond to those derived by
O'Hanlon \& Tupper~\cite{O-H+Tup 72}.
If we convert them to proper time
they read $a(t) = a_0\, t^{q_{\pm}}$ and $\phi = \phi_0\,t^{(1-3q_{\pm})}$,
where
\begin{equation}
q_{\pm} \equiv \frac{\omega}{3\left( \omega+1\pm
\sqrt{\frac{2\omega+3}{3}} \right) } \; , \label{e:qvac}
\end{equation}
[see Fig.~\ref{fig4}]. The $k\neq 0$ solutions were obtained by
Lorentz--Petzold~\cite{L-P} and by Barrow~\cite{Barrow 93}. As
$\eta\rightarrow 0$, and thus $a\rightarrow 0$ for $\omega>0$, they approach
the $k=0 $ power law behaviour. Also note that all solutions exhibit two
branches. This is a consequence of the identity $A=\pm f$ between the
integration constants. Each branch corresponds to different signs of
$\dot{\phi}/\phi$. In fact, the $q_+$ branch is associated with an increasing
$|\phi|$, which means that $G$ approaches zero in the $t\rightarrow \infty$
limit. Since this branch corresponds to a slower expansion, we shall follow
Gurevich {\it et al}~\cite{Gur+Fin+Rub 73} in calling it the {\em slow}
branch. On the contrary, the $q_-$ {\em fast} branch has a decreasing $|\phi|$
and $G$, consequently, approaches $\pm\infty$ with time.

Note that $\tau\to0$ (and thus $\eta\to0$) coincides with $a\to0$ for both
branches if and only if $\omega>0$ (and thus $\beta<1$), in agreement with our
earlier arguments. For $\omega<0$ the solutions do not have zero size at
$\tau=0$ but are still singular in the sense that the Ricci curvature scalar,
for instance, diverges.

We can choose $\phi_0$ to be either positive or negative and thus the sign of
the gravitational ``constant'' is arbitrary as we would expect for solutions
of the field equations in vacuum. Of course $a^2$ must remain positive so the
product we require $\phi_0A>0$. Because of our definition of $\tau$ in
Eq.~(\ref{etau}) we also have two distinct solutions corresponding to whether
$\tau$ decreases with time, corresponding to $\eta\leq0$ and a collapsing
universe as $\tau\to0$ for $\omega>0$, or increases with $\eta\geq0$ for a
universe expanding from $\tau=0$ if $\omega>0$.

\subsubsection*{Vacuum solution with $\omega\to\infty$}

The simplest function which includes a divergent $\omega(\phi)$ at
a finite value of $\phi=\phi_*$, is
\begin{equation}
2\omega(\phi) + 3 = \left( 2\omega_0 + 3 \right) \,
	\frac{\phi_*}{\phi_*-\phi} \; .
\end{equation}
The integral in Eq.~(\ref{eYphi}) then yields
\begin{equation}
Y(\phi) = \sqrt{\frac{2\omega_0+3}{3}}
	\, \ln\left(\frac{\sqrt{\phi_*}-\sqrt{\phi_*-\phi}}
			{\sqrt{\phi_*}+\sqrt{\phi_*-\phi}} \right)
\,  = \, \pm \, \ln{\tau} \, + {\rm constant} \; .
\end{equation}
which in turn gives
\begin{eqnarray}
\phi & = &  \phi_* \,
	\frac{4\tau_*^{\beta_0}\tau^{\beta_0}}
		{\left(\tau_*^{\beta_0}+\tau^{\beta_0}\right)^2}
\; , \\
a^2 & = & \frac{A}{\phi_*} \,
 \frac{\left(\tau_*^{\beta_0}+\tau^{\beta_0}\right)^2}
	{\tau_*^{\beta_0}\tau^{\beta_0}} \,
 \frac{\tau}{1+k\tau^2}
\; ,
\end{eqnarray}
[see Fig.~\ref{fig5}] where we have written
$\beta_0=\pm\sqrt{3/(2\omega_0+3)}$, although in fact the choice of $\pm$ is
irrelevant here for $\tau_*\neq0$. Notice again that $a\to0$ as $\tau\to0$ for
$\omega>0$. The function $\phi$ is always zero at the initial singularity
($a=0$) and increases towards its maximum value $\phi_*$ where
$\omega\to\infty$. Because $\tau$ remains bounded ($\tau\leq1$) in an open
universe, $\phi$ will never reach $\phi_*$ if $\tau_*>1$.

For $\tau>\tau_*$, $\phi$ then decreases towards zero (which it attains for
$k\ge 0$ as $\tau\to\infty$). This demonstrates that, although $\dot{\phi}\to0$
as $\phi\to\phi_*$ and $\omega\to\infty$, this is not the late time attractor
solution. Instead we require that the function $Y$ must diverge as
$\tau\to\infty$ and thus $\phi\to0$.

\subsubsection*{Vacuum solution with Brans--Dicke and G.R. limits}

Consider the function $\omega(\phi)$ such that
\begin{equation}
2\omega(\phi) +3 = \left( 2\omega_0 + 3 \right) \frac{\phi^2}{(\phi-\phi_*)^2}
\; .
\end{equation}
Clearly $\omega\to$constant as $\phi\to\infty$, but is divergent at
$\phi=\phi_*$.

Considering only $\phi>\phi_*$, initially, we have
\begin{equation}
Y(\phi) = \sqrt{\frac{2\omega_0+3}{3}} \ln \left( \frac{\phi}{\phi_*}-1 \right)
\; ,
\end{equation}
and thus
\begin{eqnarray}
\phi & = &
	\phi_* \left( 1+\left(\frac{\tau}{\tau_*}\right)^{\beta_0} \right)
\; , \\
a^2 & = & \frac{A}{\phi_*} \
 \frac{\tau_*^{\beta_0}}{\tau_*^{\beta_0}+\tau^{\beta_0}} \
 \frac{\tau}{1+k\tau^2}
\; .
\end{eqnarray}
Because $Y(\phi)$ is divergent at both $\phi=\phi_*$ and $\phi\to\infty$ we
again have two distinct branches according to whether $\dot{\phi}/\phi$ is
increasing or decreasing. In the former case, when $\beta_0>0$, we find a slow
branch where the initial general relativistic behaviour $\phi\simeq\phi_*$
turns into the Brans--Dicke solution $\phi\propto\tau^{\beta_0}$ as
$\tau\to\infty$. For $\beta_0<0$ we have decreasing $\phi$, the fast branch,
and the behaviour is reversed as $\tau\to\infty$. Note that the
$\tau\to\infty$ limit is only achieved for $k\geq0$ and that the late time
behaviour in open models always corresponds to the general relativistic
behaviour with $\phi\to\phi_*(1+\tau_*^{-\beta_0})$ as $\tau\to1$.

For $\phi<\phi_*$ we see that $2\omega+3$ may reach zero.  This allows far
more complex behaviour.
\begin{eqnarray}
\phi & = &
	\phi_* \left( 1-\left(\frac{\tau}{\tau_*}\right)^{\beta_0} \right)
\; , \\
a^2 & = & \frac{A}{\phi_*} \
 \frac{\tau_*^{\beta_0}}{\tau_*^{\beta_0}-\tau^{\beta_0}} \
 \frac{\tau}{1+k\tau^2} \; .
\end{eqnarray}
Notice now that $\phi$ reaches zero when $\tau=\tau_*$. This corresponds to
the divergence of the scale factor $a$ at a finite proper time. Thus, for
instance, the closed model does {\em not} recollapse.

\subsection{Stiff fluid solutions}

We consider in this section the case where matter is described by a barotropic
equation of state with $\gamma=2$ (stiff matter) which as we have seen
describes the long wavelength modes of a massless scalar field. In terms of the
same variables $X$ and $Y$ the field equations become
\begin{equation}
(X')^2+4kX^2- (Y'\, X)^2 = 4M\phi \; ,
\end{equation}
\begin{equation}
\left[Y'\, X\right]' = -\frac{2M\phi}{X}\,\sqrt{\frac{3}{2\omega+3}} \; ,
\end{equation}
\begin{equation}
X''+4kX=0 \; .
\end{equation}
The last equation is identical to the corresponding equation for the vacuum
case, and thus  $X(\eta)$ is given by the same expressions
(Eq.~(\ref{evacX})).  This also yields the first integral
\begin{equation}
X'^2 + 4kX^2 = \bar{A} \; ,
\end{equation}
which upon insertion into the first of the field equations leads to
\begin{equation}
Y' X = \pm \sqrt{ \bar{A} - 4M\phi}
\label{estiffm} \; .
\end{equation}
This requires that $\phi\le \bar{A} / 4M$.
Notice also that unlike the vacuum case $\bar{A}$ could be
negative, but only if $k=-1$ and $\phi$ is also
negative. This gives one extra solution for $X(\eta)$ when $k=-1$ in addition
to the vacuum solutions where $\bar{A}=-A^2$,
\begin{equation}
X(\eta) = - \frac{A}{2} \cosh 2\eta \; ,
\end{equation}
or in terms of the variable $\tau$ defined in Eq.~(\ref{etau})
\begin{equation}
X(\tau) = -A \frac{1-k\tau^2}{1+k\tau^2} \; .
\label{ensX}
\end{equation}
For $k=+1$ note that this corresponds to $X\propto\cos2\eta$ which is
equivalent simply to a different choice of the zero--point of $\eta$, but in
the open model we have a qualitatively different behaviour when $\bar{A}<0$.
$X=a^2\phi$ remains non--zero at all times and thus we can obtain
non--singular models where $a$ remains non--zero.

Now, from Eq.~(\ref{estiffm}) we derive
\begin{equation}
Y = \int \sqrt{\bar{A}-4M\phi} \frac{d\eta}{X} \; ,
\end{equation}
and thus it is useful to define
\begin{equation}
Z(\phi) \equiv \int\sqrt{\frac{2\omega+3}{3}} \,
\frac{{\rm d}\phi}{\phi\sqrt{\bar{A}-4M\phi}}=
\pm \int \,\frac{{\rm d}\eta}{X(\eta)} \; .
\label{eZphi}
\end{equation}
where the right--hand side of this equation is just $\pm\ln{\tau}$ as
for the vacuum case. Thus, just as in the vacuum case we required the function
$Y(\phi)$ to diverge as $\tau\to0$ or $\to\infty$, in the stiff fluid case
we require $Z(\phi)$ to diverge in these limits. We see that if
\begin{equation}
\frac{2\omega_{\rm vac}(\phi)+3}{A^2} =
	\frac{2\omega(\phi)+3}{\bar{A}-4M\phi}
\; ,
\label{estiffvac}
\end{equation}
the vacuum solutions for $a(t)$ and $\phi(t)$ with $\omega_{\rm vac}(\phi)$
carry over to the stiff fluid solutions for $\omega(\phi)$. The
reason for the equivalence
becomes more apparent when we discuss the conformally transformed
picture in the next section. When $\bar{A}<0$ we see that for
$2\omega+3>0$ we find the vacuum equivalent $2\omega_{\rm vac}+3<0$ which is
why we did not find the non--singular open models in the vacuum case.

The condition for $\dot{a}=0$ now becomes
\begin{equation}
\omega \ = \ \frac{6(kX^2-M\phi)}{\bar{A}-4kX^2} \; ,
\end{equation}
confirming that $\omega>0$ is compatible with a turning point for $k<0$ when
$\bar{A}<0$. For $k=0$ where we must have $\bar{A}>0$, or as $X\to0$, the
condition becomes $\omega=-6M\phi/\bar{A}$. Thus the sign of $\phi$ becomes
crucial. As one might expect, if the gravitational mass, $M/\phi$ is negative,
the initial singularity can be avoided even for $\omega>0$, while for
$M/\phi>0$ the presence of the stiff fluid requires an increasingly negative
value of $\omega$ to avoid the singularity.

\subsubsection*{Stiff fluid solution in Brans--Dicke gravity}

Proceeding as for the vacuum case,
we start by considering the $\omega=\omega_0=$constant
case which enables us to compare our results with the $k=0$ solutions
existing in the literature.

For $\bar{A} = + A^2 \ge 4M\phi$ we have
\begin{equation}
A \times Z = \sqrt{\frac{2\omega_0+3}{3}}\;\ln{\left[
\frac{A-\sqrt{A^2-4M\phi}}{A+\sqrt{A^2-4M\phi}}\right]}
	= \pm \ln{\tau} + {\rm constant} \; .
\end{equation}
Notice that this is exactly the same result as found in the vacuum case
with $2\omega_{\rm vac}(\phi)+3=(2\omega_*+3)\phi_*/(\phi_*-\phi)$ if we write
$2\omega_*+3=(2\omega_0+3)$ and $\phi_*=A^2/(4M)$.
This is a demonstration of the equivalence between different vacuum
and stiff fluid solutions given in Eq.~(\ref{estiffvac}).

Thus, for $A>0$,
\begin{eqnarray}
\phi & = & \frac{A^2}{M} \,
 \frac{\tau_*^{\beta}\tau^{\beta}}
      {\left( \tau_*^{\beta}+\tau^{\beta} \right)^2} \; , \\
a^2 & = & \frac{M}{A} \,
 \frac{\left( \tau_*^\beta+\tau^{\beta} \right)^2}
	{\tau_*^{\beta} \, \tau^{\beta}}
 \, \frac{\tau}{1+k\tau^2} \; ,
\end{eqnarray}
where $\tau_*$ is the constant of integration chosen to coincide
with the value of $\tau$ for which $\phi$ reaches its maximum possible value
$\phi_*=A^2/4M$. For $\tau>\tau_*$, $\phi$ decreases back towards
zero. [See Fig.~\ref{fig5}].

If $k=-1$, $\tau$ is bounded and will never attain
$\phi_*$ if $\tau_*>1$. In this case $\phi$ remains a
monotonically increasing function of $\tau$ approaching
$4\phi_*/(1+\tau_*^{\beta})^2$ as $\tau\to1$ and thus $t$ tends to infinity.

If on the other hand we consider $A<0$, we find a solution for $\phi<0$:
\begin{eqnarray}
\phi & = & -\frac{A^2}{M} \,
	\frac{\tau_*^{\beta}\tau^{\beta}}
		{\left(\tau_*^{\beta}-\tau^{\beta}\right)^2}
\; , \\
a^2 & = & - \frac{M}{A} \,
 \frac{\left( \tau_*^{\beta}-\tau^{\beta} \right)^2}
	{\tau_*^{\beta} \, \tau^{\beta}}
 \, \frac{\tau}{1+k\tau^2} \; .
\end{eqnarray}

It is possible to see that these solutions corresponds to the
$\omega_0>-3/2$
solution derived by Gurevich {\it et al}\cite{Gur+Fin+Rub 73} (after the
necessary translation to
their time variable; Gurevich {\it et al} use $\xi$ such that
$d\xi= d\eta/a^2$). Notice that $a=0$ at both $\tau=0$ and
$\tau=\tau_*$, demonstrating that a turning point can indeed occur for
$\omega>0$ even in open or flat models in the presence of the stiff fluid if
$\phi<0$. As $\tau$ approaches $\tau_*$ from below the solution approaches
Nariai's power law solution\cite{Nariai 68}, but it is clear that this is not
the late time behaviour suggested by Gurevich {\it et al} but rather a
recollapse at a finite proper time. The correct late time behaviour for
expanding $k=0$ models is where they approach the vacuum solution as
$\tau\to\infty$ with $\phi$ positive or negative.

When $\bar{A}=-A^2$, possible only for $k<0$, we have $X(\eta)$ given by
Eq.~(\ref{ensX}) and
\begin{equation}
A \times Z(\phi) = 2 \ \sqrt{\frac{2\omega+3}{3}} \
 \tan^{-1} \left(\frac{\sqrt{-4M\phi-A^2}}{A}\right)
 = \pm 2 \ \tan^{-1}\tau + {\rm constant} \; .
\end{equation}
This gives
\begin{equation}
\phi = - \frac{A^2}{4M} \sec^2 \left( c + \beta \tan^{-1}\tau \right)
\; .
\end{equation}
Thus $\phi\leq-A^2/4M$ as required. However $\phi\to-\infty$ whenever
$\tau=\tan((\pi/2-c)/\beta)$ leading to $a=\sqrt{X/\phi}\to0$. This can always
occur when the arbitrary constant $c$ is sufficiently close to $\pi/2$
regardless of the sign of $k$.

\subsubsection*{Stiff fluid solution with Brans--Dicke and G.R. limits}

If we consider again the function
$2\omega(\phi)+3=(2\omega_0+3)\phi^2/(\phi-\phi_*)^2$ this time in the
presence of a stiff fluid, we can integrate Eq.~(\ref{eZphi}) for
$\phi>\phi_*$ to give
\begin{equation}
Z(\phi) = \sqrt{\frac{2\omega_0+3}{3}} \ \frac{1}{\sqrt{A^2-4M\phi_*}}
 \ \ln \left| \frac{\sqrt{A^2-4M\phi}-\sqrt{A^2-4M\phi_*}}
		{\sqrt{A^2-4M\phi}+\sqrt{A^2-4M\phi_*}} \right|
\; .
\end{equation}
Here $\phi$ must be constrained to lie within $\phi_*<\phi<A^2/4M$ and
so can never reach the asymptotic Brans--Dicke limit as $\phi\to\infty$.
We find
\begin{equation}
\phi = \frac{ (\tau_*^B-\tau^B)^2\phi_* + \tau_*^B\tau^B (A^2/M) }
		{(\tau_*^B+\tau^B)^2}
\; ,
\end{equation}
where we have written
\begin{equation}
B = \sqrt{ \frac{A^2-4M\phi_*}{A^2} \ \frac{3}{2\omega_0+3} } < \beta_0 \; .
\end{equation}
Thus $\phi=\phi_*$ at $\tau=0$, and reaches a maximum of $\phi=A^2/4M$ when
$\tau=\tau_*$ (possible only for $\tau_*<1$ in the open model). At late times
for $k\geq0$, as $\tau\to\infty$, $\phi$ returns to the general relativistic
result, $\phi\to\phi_*$, $\omega\to0$.

Once again for $\phi<\phi_*$ we find a considerably more complicated behaviour
where we may have $\phi\to0$ for non--zero $X$.

\subsection{Radiation solutions}
\label{srad}

The other case in which the non--vacuum equations of motion simplify
considerably is where the energy--momentum tensor is traceless ($\gamma=4/3$),
i.e.~a radiation fluid corresponding to the short wavelength modes of a
massless field.
As this case has been discussed elsewhere\cite{Barrow 93} we will describe
the behaviour only briefly for comparison with the stiff fluid case, while
presenting our results in a more compact form in terms of the time coordinate
$\tau(\eta)$.

The field equations in the presence of radiation with density
$\rho=3\Gamma/8\pi a^4$ where $\Gamma$ is a constant, become
\begin{eqnarray}
(X')^2 + 4kX^2 - (Y'X)^2 = 4\Gamma X \; , \\
\label{eRADconstraint}
(Y'X)' = 0 \; , \\
X'' + 4kX = 2\Gamma \; .
\label{eXrad}
\end{eqnarray}
The final equation can again be integrated directly to give the first equation
where $(Y'X)^2=A^2=$constant. Notice that unlike the stiff fluid case this
constant cannot be negative. The general solution of the equation of motion
for $X$ is then
\begin{equation}
X = \frac{\tau(A+\Gamma\tau)}{1+k\tau^2} \; ,
\end{equation}
in terms of the time coordinate $\tau(\eta)$ introduced in Eq.~(\ref{etau}).

The Brans--Dicke field is not driven by matter and we have the same integral
for $Y(\phi)$ as in the vacuum case, although we have a different $X(\eta)$:
\begin{equation}
Y(\phi) = \pm \int \frac{Ad\eta}{X} =
 \pm \ln\left|\frac{\Gamma\tau}{A+\Gamma\tau}\right|
 + {\rm constant} \; .
\end{equation}
The evolution of $\phi(\eta)$ is thus the same as for the vacuum case
if we replace the function $\tau(\eta)$ by
\begin{equation}
s(\eta) = \left| \frac{\Gamma\tau(\eta)}{A+\Gamma\tau(\eta)} \right| \; .
\end{equation}
Note that in spatially flat or closed models as $\tau\to\infty$ we find
$s\to1$, i.~e.~$\phi$ approaches a fixed value. In open models as
$\tau\to1$ we have $s\to\Gamma/(A+\Gamma)$. On the other hand at early
times these solutions approach the vacuum solutions as
$s\simeq(\Gamma/A)\tau$ amounts simply to a rescaling of the conformal
time or, equivalently, the scale factor.

It is now straightforward to write down the radiation solutions for the
particular choices of $\omega(\phi)$ given in the vacuum and stiff fluid cases.

The value of the $\omega$ at any turning point is now given by
\begin{eqnarray}
\omega & = & \frac{6(kX^2-\Gamma X)}{A^2-4(kX^2 - \Gamma X)} \; , \\
& = & - \left( \frac{6\Gamma X}{A^2+4\Gamma X} \right)
	\quad {\rm for}\; k=0 \; {\rm or} \; X\to0 \; .
\end{eqnarray}
The denominator must always be positive (by Eq.~(\ref{eRADconstraint})) and
thus we find again that to obtain a turning point with $\omega>0$ requires
either $k>0$, which corresponds to the usual recollapse in closed models, or
$X$ (and thus $\phi$) negative.

\subsubsection{Radiation solution in Brans--Dicke gravity}

\begin{eqnarray}
\phi & = & \phi_* \ s^{\pm\beta} \; , \\
a^2 & = & \frac{1}{\phi_*} \
 \frac{s^{\mp\beta}\tau(A+\Gamma\tau)}{1+k\tau^2} \; .
\end{eqnarray}
As is well known, this Brans--Dicke solution approaches the general
relativistic solution with constant $\omega$ at late times during the
radiation dominated era.

\subsubsection{Radiation solution with Brans--Dicke and G.R. limits}

When $2\omega(\phi)+3=(2\omega_0+3)\phi^2/(\phi-\phi_*)^2$ we find
\begin{eqnarray}
\phi & = & \phi_* \ \left( 1+\left(\frac{s}{s_*}\right)^{\beta_0}
				 \right) \; , \\
a^2 & = & \frac{1}{\phi_*} \ \frac{s_*^{\beta_0}}
					{s_*^{\beta_0}+s^{\beta_0}}
 \ \frac{\tau(A+\Gamma\tau)}{1+k\tau^2} \; .
\end{eqnarray}
Once again $\phi$ approaches a constant at late times however, unlike the
stiff fluid case considered earlier, this constant value may not be close to
$\phi_*$ so this need not coincide with $\omega\to\infty$.

\section{Conformally transformed frame}
\label{sectCONF}

It has long been realised that a theory with varying gravitational
coupling such as scalar--tensor gravity must be equivalent to one in
which the gravitational coupling is constant but masses and lengths
vary~\cite{Dicke 62}.
Mathematically this equivalence can be shown by using a conformally
rescaled metric
\begin{equation}
\tilde{g}_{ab} = \left( \frac{\phi}{\phi_0} \right) g_{ab} \; .
\end{equation}
$\phi_0$ is just an arbitrary constant introduced to keep the conformal
factor dimensionless.
Written in terms of this new metric and its scalar curvature,
$\tilde{R}$, the scalar--tensor action given in Eq.~(\ref{eSTTaction})
becomes
\begin{equation}
S = \frac{1}{16\pi} \int_M d^4x \sqrt{-\tilde{g}} \left[
	\phi_0 \tilde{R} - 16\pi \left(
		- \frac{1}{2}\tilde{g}^{ab}\psi_{,a}\psi_{,b}
		+ \left(\frac{\phi_0}{\phi}\right)^2 {\cal L}_{\rm matter}
	\right) \right] \; ,
\end{equation}
where we introduce a new scalar field $\psi(\phi)$ defined by
\begin{equation}
d\psi \equiv
	\sqrt{\phi_0 \, \frac{2\omega+3}{16\pi}} \, \frac{d\phi}{\phi} \; .
\label{defpsi}
\end{equation}
The gravitational Lagrangian is reduced simply to the Einstein--Hilbert
Lagrangian of general relativity, albeit at the expense of altering the matter
Lagrangian. Thus we shall refer to this as the Einstein frame.

The arbitrary dimensional constant $\phi_0$ plays the role of Newton's
constant, $G\equiv\phi_0^{-1}$. In order to avoid changing the signature, the
conformal factor relating the metrics must be positive. So for $\phi<0$ we
must pick $\phi_0<0$ giving a negative gravitational constant in the Einstein
frame. Not surprisingly then, the usual singularitiy theorems need not apply
even in the Einstein frame for $\phi<0$. Similarly, in the definition of $\psi$
we require $\phi_0(2\omega+3)>0$. If this were not the case we could instead
define a scalar field
\begin{equation}
d\bar{\psi} \equiv
	\sqrt{- \, \phi_0 \, \frac{2\omega+3}{16\pi}} \, \frac{d\phi}{\phi} \; ,
\end{equation}
but this would have a negative kinetic energy density, again invalidating
the usual singularity theorems by breaking the dominant energy condition.
However for $\phi>0$ and $\omega>-3/2$ the FRW models must contain
singularities in the conformal Einstein frame where $\tilde{a}\to0$.

\subsection{Vacuum solutions}

The field equations are then, at least in vacuum (${\cal L}_{\rm
matter}=0$), just the usual Einstein field equations of general relativity
plus a massless scalar field, $\psi$. In particular, in a FRW universe
(which remains homogeneous and isotropic under the homogeneous
transformation) we have
\begin{eqnarray}
\tilde{H}^2 = \frac{8\pi}{3} \, \frac{\hat{\rho}}{\phi_0} \, - \,
	\frac{k}{\tilde{a}^2} \; ,\\
\frac{d^2\psi}{d\tilde{t}^2} + 3\tilde{H} \frac{d\psi}{d\tilde{t}} = 0
 \; ,\\
\frac{d\tilde{H}}{d\tilde{t}} + \tilde{H}^2 =
 - \frac{4\pi}{3} \frac{\hat{\rho}+3\hat{p}}{\phi_0} \; ,
\end{eqnarray}
where the scale factor in the conformal frame
$\tilde{a}=(\phi/\phi_0)^{1/2}a$, $d\tilde{t}=(\phi/\phi_0)^{1/2}dt$ and
$\tilde{H}=(d\tilde{a}/d\tilde{t})/\tilde{a}$. (Note that $\tilde{t}$ is the
time in the conformal frame and not to be confused with the conformally
invariant time, $\eta$, used earlier.)
The massless scalar field behaves, as it
must, as a stiff fluid with density
$\hat{\rho}=\hat{p}=(d\psi/d\tilde{t})^2/2$.

Notice now that the variables $X$ and $Y$ introduced in the previous
section correspond to the square of the conformal scale factor and the
scalar field $\psi$ respectively.
\begin{eqnarray}
X & \equiv & \frac{a^2}{\phi} \equiv \frac{\tilde{a}^2}{\phi_0} \; , \\
Y & \equiv & \int \, \sqrt{\frac{2\omega+3}{3}} \, \frac{d\phi}{\phi}
	\equiv \sqrt{\frac{16\pi}{3\phi_0}} \, \psi \; .
\label{eYphipsi}
\end{eqnarray}
The equations of motion for the conformal scale factor written in terms of $X$
and for $\psi$ written in terms of $Y$ and derivatives with respect to the
conformal time $\eta$ are then precisely Eqs.~(\ref{eLPa}--\ref{eLPc}) solved
in Section~\ref{sVac}.

We can solve explicitly for $X$ and $Y$ as functions of $\eta$ because the
stiff fluid continuity equation can be integrated directly (as for any perfect
barotropic fluid) to give $\hat{\rho}\propto\tilde{a}^{-6}$. These results are
independent of the form of $\omega(\phi)$. A particular choice of
$\omega(\phi)$ determines how $\phi$ is related to the stiff fluid field
$\psi$. To obtain $\phi(\eta)$ we must be able to perform the integral in
Eq.~(\ref{eYphipsi}), and thus we also obtain the scale factor in the original
frame, $a\equiv (\phi_0/\phi)^{1/2}\tilde{a}$.

\subsection{Non--vacuum solutions}

If we include the matter lagrangian for a perfect fluid in the
original scalar--tensor frame, then there is a non--trivial interaction
between this matter and the scalar field, $\psi$, in the Einstein frame.
\begin{equation}
\tilde{\cal L}_{\rm matter} = \left( \frac{\phi_0}{\phi(\psi)} \right)^2
	{\cal L}_{\rm matter} \; .
\end{equation}
Thus the matter energy--momentum tensor, defined in the Einstein metric,
\begin{equation}
\tilde{T}^{ab} \equiv \frac{2}{\sqrt{-\tilde{g}}}
	\frac{\partial}{\partial\tilde{g}_{ab}} \left(
	\sqrt{-\tilde{g}} \tilde{\cal L}_{\rm matter} \right) \; ,
\end{equation}
is no longer independently conserved,
\begin{equation}
\tilde{\nabla}^a \tilde{T}_{ab} =
	- \frac{1}{2\sqrt{\phi_0}}
		\sqrt{\frac{16\pi}{2\omega+3}} \tilde{T}^a_a \psi_{,b}
\; ,
\end{equation}
unless it is traceless, i.e. vacuum or radiation.
The conformally transformed density, $\tilde{\rho}=(\phi_0/\phi)^2\rho$, and
pressure, $\tilde{p}=(\phi_0/\phi)^2p$, of the fluid become dependent
on $\phi$ and thus $\psi$, so while the fluid retains the same barotropic
equation of state it is no longer a perfect fluid in general.
Note however that the
overall energy--momentum tensor of the matter plus the $\psi$ field must
be conserved as guaranteed in general relativity by the Ricci identity.

We have the usual general relativistic equations of motion in a FRW model
\begin{eqnarray}
\tilde{H}^2 = \frac{8\pi}{3} \, \frac{\hat{\rho}+\tilde{\rho}}{\phi_0} \, - \,
	\frac{k}{\tilde{a}^2} \; ,\\
\frac{d\tilde{H}}{d\tilde{t}} + \tilde{H}^2 =
 - \frac{4\pi}{3} \, \frac{\hat{\rho}+\tilde{\rho}+3(\hat{p}+\tilde{p})}
			{\phi_0} \; ,
\end{eqnarray}
and the interaction leads to a transfer of energy between the original fluid
and the stiff ($\psi$) fluid:
\begin{eqnarray}
\frac{d\tilde{\rho}}{d\tilde{t}} & = & -3\tilde{H}(\tilde{\rho}+\tilde{p})
	+\frac{1}{2\sqrt{\phi_0}} \sqrt{\frac{16\pi}{2\omega+3}}
		(3\tilde{p}-\tilde{\rho}) \frac{d\psi}{d\tilde{t}} \; , \\
\frac{d\hat{\rho}}{d\tilde{t}} & = & -6\tilde{H}\hat{\rho}
	-\frac{1}{2\sqrt{\phi_0}} \sqrt{\frac{16\pi}{2\omega+3}}
		(3\tilde{p}-\tilde{\rho}) \frac{d\psi}{d\tilde{t}}
\; .
\end{eqnarray}

Again we find two cases in which the problem simplifies. Firstly for radiation
($\tilde{\rho}=3\tilde{p}$) there is no interaction and both continuity
equations can be directly integrated and the conformal picture contains two
non--interacting fluids:
\begin{eqnarray}
\frac{8\pi}{3\phi_0} \ \hat{\rho} & = & \frac{A^2}{4\tilde{a}^6} \; , \\
\frac{8\pi}{3\phi_0} \ \tilde{\rho}_{\rm rad} & = &
	\frac{\Gamma}{\tilde{a}^4} \; .
\end{eqnarray}
This is precisely the case considered recently by Barrow~\cite{Barrow
93} (although without explicitly invoking the conformal frame) and
discussed in Section~\ref{srad}.

The second case in which we can find exact
solutions is where the original fluid is itself a stiff fluid (or
massless scalar field) in which case although there is an interaction
between the two fluids, their combined dynamical effect is that of a
single perfect stiff fluid\footnote{
The combined energy--momentum tensor of two interacting fluids is equivalent
to that of a single perfect fluid provided their velocity fields are parallel.
This must be true if both fluids are homogeneous as is the case here.
Futhermore as they are both stiff fluids, $p=\rho$, in this case, their total
pressure must be equal to their total density.
}, or massless scalar field $\chi$ say:
\begin{equation}
\frac{8\pi}{3\phi_0} \, \tilde{\rho}_{\chi} =
 \frac{8\pi}{3\phi_0} \left( \tilde{\rho} + \hat{\rho} \right)
	= \frac{\bar{A}}{4\tilde{a}^6}
\; .
\end{equation}
This is why we find exactly the same equation of motion for the scale
factor in the conformal frame, $\tilde{a}^2\propto X$, in the stiff fluid
case as in the vacuum case. Notice now that in the conformal frame we
must have $\bar{A}=+A^2>0$ for a positive energy density. The non--singular
solutions found when $k<0$ and $\bar{A}<0$ with a stiff fluid in the Jordan
frame correspond to solutions with negative enrgy density in the Einstein
frame.

The continuity equation for the original fluid can always be integrated
to give
\begin{equation}
\frac{8\pi}{3\phi_0} \, \tilde{\rho} = \frac{M}{\tilde{a}^{3\gamma}}
	\left(\frac{\phi_0}{\phi}\right)^{(4-3\gamma)/2}
\; ,
\end{equation}
and so in the stiff fluid case we have
\begin{equation}
\frac{8\pi}{3\phi_0} \, \hat{\rho} =
	\frac{4\pi}{3\phi_0} \, \left( \frac{d\psi}{d\tilde{t}} \right)^2
	= \frac{A^2 - 4M\phi}{4\tilde{a}^6} \; .
\label{erhohat}
\end{equation}
We have $\tilde{a}^2\propto X$ as a function of $\eta$ and we must now
perform the integral in Eq.~(\ref{eZphi}) to obtain $\phi(\eta)$. The
change in the relation between $\phi$ and the total stiff fluid density in the
Einstein frame compared with the vacuum case is equivalent to a
different choice of $\omega(\phi)$ (which relates $\phi$ to $\psi$)
as demonstrated in Eq.~(\ref{estiffvac}).
The vacuum case can of course be seen as a special case amongst the
stiff fluid solutions, where $M=0$, and thus
$\omega(\phi)=\omega_{\rm vac}(\phi)$.

We can also obtain exact solutions for radiation {\em and} stiff
fluid in the original Jordan frame as the radiation remains decoupled in the
Einstein frame and the interaction is solely between the two stiff
fluids in that frame. Thus the equation of motion for the scale factor in the
conformal frame is exactly the same as in
the radiation only case,
Eq.~(\ref{eXrad}), while the equation for $\phi$ is the same as in
the stiff fluid case, Eq.~(\ref{eZphi}).

\subsubsection*{Stiff fluid plus radiation in Brans--Dicke gravity}

To solve for the evolution of Brans--Dicke models (where $\omega_0=$constant)
in the presence of both radiation and a stiff fluid,  the conformal frame is
particularly useful as the evolution of the conformal scale factor, or
$X\equiv\tilde{a}^2/\phi_0$, is exactly the same as for radiation only
(Eq.~(\ref{eXrad})). The evolution of $\phi$ then follows directly from
Eq.~(\ref{erhohat}) as
\begin{eqnarray}
\hat{\rho} & = & \frac{3\phi_0}{32\pi X} \, \left(\frac{2\omega+3}{3}\right)
 \left(\frac{1}{\phi}\frac{d\phi}{d\eta}\right)^2 \; , \nonumber \\
& = & \frac{3\phi_0}{32\pi} \, \frac{A^2-4M\phi}{X^3} \; ,
\end{eqnarray}
so that, from the definition of $\psi$ in Eq.~(\ref{defpsi}),
\begin{equation}
\sqrt{\frac{2\omega_0+3}{3}} \
 \ln\left[ \frac{A-\sqrt{A^2-4M\phi}}{A+\sqrt{A^2-4M\phi}} \right]
= \pm \ln\left|\frac{\Gamma\tau}{A+\Gamma\tau}\right| + {\rm constant} \; .
\end{equation}
Rewriting this to give $\phi$ and thus $a=\sqrt{X/\phi}$ yields
\begin{eqnarray}
\phi & = & \frac{A^2}{M} \,
 \frac{s_*^\beta s^\beta}{(s_*^\beta +s^\beta)^2} \; , \\
a^2 & = & \frac{M}{A^2} \frac{(s_*^\beta+s^\beta)^2}
				{s_*^\beta s^\beta}
 \frac{\tau (A+\Gamma\tau)}{1+k\tau^2} \; ,
\end{eqnarray}
where
\begin{equation}
s(\eta) = \left| \frac{\Gamma\tau}{A+\Gamma\tau} \right| \; ,
\end{equation}
$s_*$ is a constant of integration and $\beta=\sqrt{3/(2\omega_0+3)}$. Thus the
behaviour is very similar to that seen for stiff fluid only, except that the
variable $s$ takes the place of $\tau$.

Unlike $\tau$, $s\to$constant at late times for $k=0$ (where $s\to1$) as well
as for $k<0$ (where $s\to\Gamma/(A+\Gamma)$). Thus the Brans--Dicke field
becomes frozen in at late times in the flat FRW universe, dominated by the
friction due to Hubble expansion driven by radiation, just as it is in the
open FRW model where the expansion becomes driven by the curvature. Only in
the closed universe does the dynamical effect of the stiff fluid remain
important. Note however that the radiation delays the recollapse which occurs
at $\eta>\pi/2$. This means that $\tau\equiv \tan(\eta)$ becomes negative, but
the solution is still well behaved as $s>0$ and both the conformal Einstein
and Jordan (for $\omega>0$) scale factors recollapse, $X$,$a\to0$, when
$\eta=\pi+\tan^{-1}(-A/\Gamma)$, as $s\to\infty$.

Notice once again that the presence of a stiff fluid in the Jordan frame just
leads to solutions which would be obtained in the absence of the stiff fluid
but with the modified $\omega_{\rm vac}(\phi)$ given in Eq.~(\ref{estiffvac}).

\section{Conclusions}

We have shown how to extend the procedure recently proposed by
Barrow\cite{Barrow 93} to obtain the solutions for general $\omega(\phi)$
scalar--tensor gravity theories with a stiff fluid in addition to vacuum or
radiation solutions in a FRW metric. These two non--vacuum cases correspond
to the extreme long and short wavelength modes respectively of a minimally
coupled massless scalar field. We show that these solutions can be obtained
due to the particularly simple evolution of the corresponding scale factor in
the conformally related Einstein frame which is independent of the form of
$\omega(\phi)$.

In the presence of a stiff fluid the scale factor evolves like a
scalar--tensor cosmology with a modified $\omega(\phi)\to\omega_{\rm
vac}(\phi)$, as defined in Eq.~(\ref{estiffvac}). This is because introducing
a new scalar field modifies the relation between $\phi$ and the total energy
density in the homogeneous scalar fluid. For example the Brans--Dicke model
($\omega=$constant) in the presence of a stiff fluid evolves like a vacuum
model with $\omega_{\rm vac}(\phi) \propto \phi_*/(\phi_*-\phi)$. This
significantly modifies the evolution of the Brans--Dicke field leading to an
upper bound of $\phi\leq\phi_*$.

We find that even for functions $\omega(\phi)$ that diverge at a finite
value of $\phi$, this need not be a stable late time attractor for $k=0$
models, in contrast to Damour and Nordvedt's
rule\cite{Damour+Nordvedt 93}
that $\omega\to\infty$ is a cosmological attractor. Instead (due to the absence
of the damping effect of matter with $p<\rho$, required by Damour and
Nordvedt's
result) we find that the late (or early) time attractor in vacuum, as
$a\to\infty$ (or $a\to0$), is
associated with the divergence of the function
$Y(\phi)\propto\int\sqrt{2\omega+3}d\phi/\phi$.
In the presence of a stiff fluid the function $Z(\phi)\propto
\int(\sqrt{(2\omega+3)/(A^2-4M\phi)}d\phi/\phi$ must diverge
as $a\to0$ or $\to\infty$.

The stiff fluid solutions are expected to be of primary importance as the scale
factor $a\to0$. When spatial curvature is negligible ($k=0$), the condition
necessary for a turning point, $\dot{a}=0$, in the stiff fluid cosmology is
simply $\omega=-6M\phi/\bar{A}$ where $\bar{A}$ and $M$ are positive constants
of integration. In vacuum this reduces to $\omega=0$ and the sign of $\phi$ is
irrelevant. In the conformally related Einstein frame we have seen that the
evolution is simply that for a stiff fluid irrespective of the form of
$\omega(\phi)$ and thus singularities are always present here provided
$\omega>-3/2$. Only for $\omega<-3/2$ does the energy density of the stiff
fluid in the Einstein frame become negative and so non--singular behaviour
becomes possible.

\section*{Acknowledgements}

DW is supported by the SERC. The authors are grateful to John Barrow and Ed
Copeland for useful discussions, and acknowledge use of the Starlink computer
system at Sussex.




\section*{Figure captions}

\begin{figure}
\caption{The function $X$, defined in Eq.~(\protect\ref{eg0}) plotted
against conformal time, $eta$. The solid line represents $k=0$, the dotted line
$k=+1$ and the short dashed line $k=-1$ models. The long dashed line is the
non--singular function for $k=-1$ given in Eq.(\protect\ref{ensX}).}
\label{fig1}
\end{figure}

\begin{figure}
\caption{Vacuum solutions for Brans--Dicke cosmologies showing Brans--Dicke
field $\phi$ and scale factor $a$ against proper time in the Jordan frame for
the {\it fast} branch where $\phi=0$ at $a=0$. Again the solid line represents
$k=0$, the dotted line $k=+1$ and the dashed line $k=-1$ models.}
\label{fig2}
\end{figure}

\begin{figure}
\caption{Same as Fig.~\protect\ref{fig2} but showing the {\it slow} branch.}
\label{fig3}
\end{figure}

\begin{figure}
\caption{Graph showing the exponents, $q_-(\omega)$ (the fast branch) and
$q_+(\omega)$ (the slow branch), of power law expansion for $k=0$ vacuum
Brans--Dicke cosmologies.}
\label{fig4}
\end{figure}

\begin{figure}
\caption{Vacuum solutions for scalar--tensor gravity theory with
$2\omega+3=9\phi_*/(\phi_*-\phi)$ showing Brans--Dicke field and scale factor
against proper time in the Jordan frame. Note that $\omega\to\infty$ is not
a late time attractor. This is identical to the evolution of a Brans--Dicke
model in the presence of a stiff fluid.}
\label{fig5}
\end{figure}

\end{document}